\documentclass[a4paper,12pt]{article}
\usepackage[english]{babel}
\usepackage{epsfig,amsmath}
\usepackage{cite}
\date{}
\begin{document}
\title{
\textbf{Comparative study of mixed monolayers of a calixarene with a fatty
acid and a non-amphiphilic nematic liquid crystal}} \normalsize
\author{
F. Nannelli$^{1}$, V. S. U. Fazio$^{1*}$,\\ Y. Matsuzawa$^{2}$,
L. Komitov$^{1}$, K. Ichimura$^{2}$\\
\small
$^{1}$ Department of Microelectronics and Nanoscience,
Liquid Crystal Physics,\\
\small
Chalmers University of Technology \& G\"oteborg University, S-41296
\small
G\"oteborg, Sweden\\
\small
$^{2}$ Research Laboratory of Resources Utilization,
Tokyo Institute of Technology,\\ 
\small
Nagatsuta, Midori-ku, Yokohama 226-8303, Japan \normalsize
}
\maketitle

\noindent
\textbf{Abstract} \,\,\,\,\,\,\, 
Many nematic liquid crystals (NLCs) lack a polar headgroup and thus 
are not able to form stable monolayers at the air/water (a/w) interface.
A way to obtain monomolecular films of these compounds is to 
incorporate them into host monolayers of amphiphilic molecules.
We report a comparative investigation of mixed films of 
Calix[4]resorcinarene O-octacarboxymethylated (CRO) with the 
non-amphiphilic NLC MBBA and with the amphiphilic stearic acid (SA).
The comparative study is useful for a better understanding of the 
characteristics of the CRO-MBBA mixed films.
Surface-pressure and surface-potential measurements on the monolayers 
at a/w interface, as well as ellipsometric and spectroscopic measurements 
on transferred Langmuir-Blodgett (LB) films, confirm that MBBA is 
additively incorporated into CRO films.

\vspace{6mm}

\section{Introduction}
Monolayers at the air/water (a/w) interface are well-defined systems 
that are used to study the surface properties of amphiphilic 
compounds\cite{Gaines}.
Such monolayers can be transferred onto solid substrates by the 
Langmuir-Blodgett (LB) technique, which ensures the deposition of 
mono- or multi-layered films with controlled molecular order and 
thickness in addition to a very high reproducibility\cite{Petty}.
The lack of a hydrophilic headgroup usually disenables a molecule 
in forming organized monolayers at the a/w interface.
However, in some cases such molecules have been incorporated into 
monolayers of amphiphilic compounds \cite{SchTewKuh74, CorMob92}.

MBBA is a hydrophobic non-amphiphilic molecule and does not form 
ordered monolayers at the a/w interface\cite{FazKomLagMobXX}.
Therefore, no LB layers of this compound can be deposited.
Calix[4]resorcinarenes are cyclic olygomers made of benzene rings 
which form polar cavities where guest molecules can be trapped.
They have been extensively studied and a number of 
significant industrial applications have been reported\cite{Gutsche}.
Current interest in possible applications of calixresorcinarenes 
includes phenomena such as binding\cite{IchFukFujKaw97} and 
incorporation\cite{MatSekIch98} of guest compounds.

The aim of this work is to understand the interactions between CRO 
(host molecule) and MBBA (non-amphiphilic guest compound) in films at
the a/w interface and in LB films making a comparative study of CRO-MBBA 
mixed monolayers with mixed monolayers of CRO and an amphiphilic compound.
As amphiphile, stearic acid (SA) was chosen because of the simple 
structure, with a carboxylic hydrophilic head and one aliphatic chain.
Monolayers at the a/w interface and LB films of this compound 
have been widely studied (see for instance \cite{Ulman}).
It was necessary first to establish the conditions of formation of 
monolayers and LB films of CRO, and then to compare these results with 
those from the guest-host systems in which new interactions may occur.

\section{Experiment}
\begin{figure}
\begin{center}
\epsfig{file=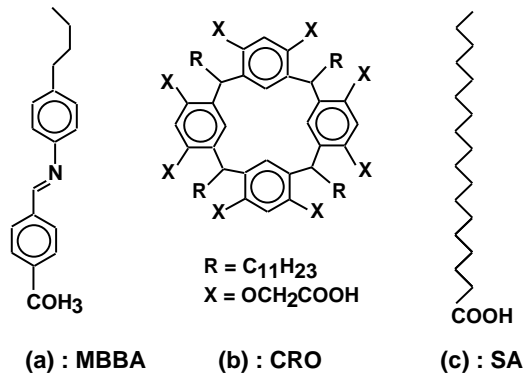, width=0.5\textwidth}
\caption{
\label{molecole}
\small{
Compounds used in this work.
(a) Nematic liquid crystal $N$-(4-methoxybenzylidene)-4-butylaniline 
(MBBA).
(b) Calix[4]resorcinarene O-octacarboxymethylated (CRO).
(c) Stearic acid (SA).
}}
\end{center}
\end{figure}

In Figure \ref{molecole} the structures of MBBA, CRO, and SA are shown.

The properties of the single and mixed monolayers at the a/w
interface have been studied by recording the surface-pressure/area 
($\pi/A$, KSV5000 trough) and the surface-potential/area ($\Delta 
V/A$, ionizing electrode method\cite{Petty} using an $^{241}$Am 
electrode\cite{GabNicDei86}) isotherms.
The substances were spread from their chloroform solutions (0.05\,mM) 
and the solvent was allowed to evaporate during 10\,min before 
compression.
Ultrapure Milli-Q water (resistivity $>$ 18\,M$\Omega$, pH = 5.5) was 
used as subphase solution.

LB films were prepared transferring layers onto sputtered 
chromium and quartz substrates (deposition pressure 25\,mN\,m$^{-1}$, 
extraction speed 3\,mm\,min$^{-1}$, immersion speed 
5\,mm\,min$^{-1}$) for ellipsometric (Rudolph Research Ellipsometer)
and spectroscopic (Bio-Rad spectrophotometer) measurements, respectively.
Transfer quality was followed by inspection of the transfer ratio.

\section{Results and discussion}

\subsection{Monolayers at the air/water interface}

\begin{figure}
\begin{center}
\begin{minipage}{\textwidth}
\parbox{0.4\textwidth}{
\epsfig{file=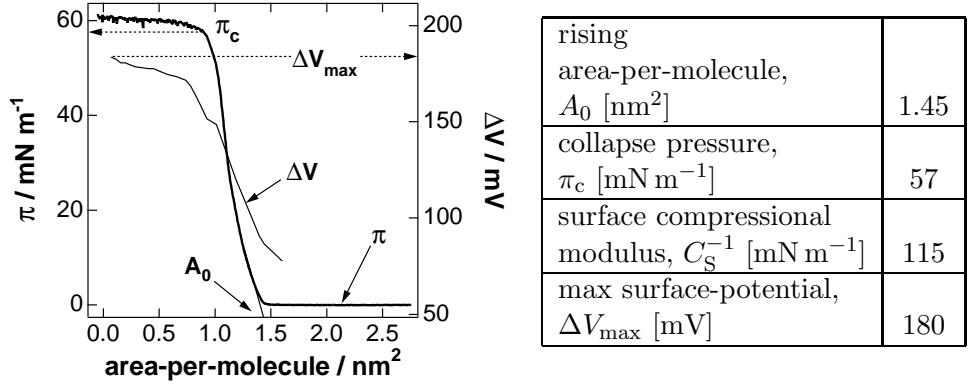, width=0.5\textwidth}
}\hfill
\parbox{0.47\textwidth}{
\small
\begin{tabular}{|l|c|}
\hline
rising & \\
area-per-molecule,  &  \\
$A_{0}$ [nm$^{2}$] & 1.45 \\
\hline
collapse pressure, & \\
$\pi_{\text{c}}$ [mN\,m$^{-1}$] & 57 \\
\hline
surface compressional  & \\
modulus, $C_{\text{S}}^{-1}$ [mN\,m$^{-1}$] & 115 \\
\hline
max surface-potential, & \\
$\Delta V_{\text{max}}$ [mV] & 180 \\
\hline
\end{tabular}
\normalsize
}
\end{minipage}
\caption{
\label{calix-ispot_a}
\small{
$\pi/A$ and and $\Delta V/A$ isotherms of pure CRO from chloroform 
solution at T=20$^{\circ}$.
Some important monolayer parameters are listed in the table.}}
\end{center}
\end{figure}

\noindent
In general, the incorporation or the binding of guest-molecules 
into the monolayer of a host molecule is ruled by the physico-chemical 
properties of the host monolayer.
Parameters such as surface-concentration, surface-potential, and 
orientation of the CRO molecules at the a/w interface are expected to 
play an important role in the host-guest interaction.
In Figure \ref{calix-ispot_a} the $\pi/A$ and $\Delta V/A$ isotherms of 
pure CRO from chloroform solution are shown together with some monolayer 
parameters extracted from the isotherms.
In general $\pi/A$ isotherms of calixresorcinarenes are influenced by 
the nature of the spreading solvent \cite{MoeDutAro94} because of the 
inclusion of solvent molecules into the CRO cavity.
In other words, the solvent has an effect on the molecular packing of 
CRO at the a/w interface.
In this context it was important to ensure the reproducibility of the 
isotherms.
CRO monolayers, with chloroform as spreading solvent, were found to 
be reproducible and stable, since no surface aggregation at the 
interface and no solubilization in the water subphase occurred during 
compression: isotherms made from spreading solutions of different 
molarities were identical.

The rising area-per-molecule, $A_{0}$, is consistent with the presence 
of eight COOH groups, each of which occupies approximately 0.18\,nm$^{2}$.
No change in the liming area with temperature was observed. 
The compressional modulus $C_{\text{S}}^{-1}$ is defined as
(see for instance \cite{Gaines, Petty}):
\begin{equation}
C_{\text{S}}^{-1} = A \left( \frac{\partial \pi}{\partial A} 
\right)_{T},
\end{equation}
where $A$ is the area-per-molecule.
The value of 115\,mN\,m$^{-1}$ indicates that the film is in a 
liquid--no-strictly-condensed phase in the entire field of existence
of the monolayer. 

From the surface-potential measurements it was possible to 
estimate the tilt angle of the CRO chains.
$\Delta V_{\text{max}}$ is proportional to the vertical component of 
the molecular dipole moment $\mu$: 
\begin{equation}
\Delta V_{\text{max}} = \frac{\mu \cos\theta}{A\,\varepsilon_{0}}.
\end{equation}
We obtained $\mu_{\text{exp}} =$ 0.4\,D from the CRO $\Delta V/A$ 
isotherm at 25\,mN\,m$^{-1}$ (LB deposition pressure) which, compared 
to the estimated value of 0.6\,D obtained from semi-empirical calculations 
using th PM3 method\cite{Ste89}, gave a tilt angle of 48 degrees.

\begin{figure}[t]
\begin{center}
\epsfig{file=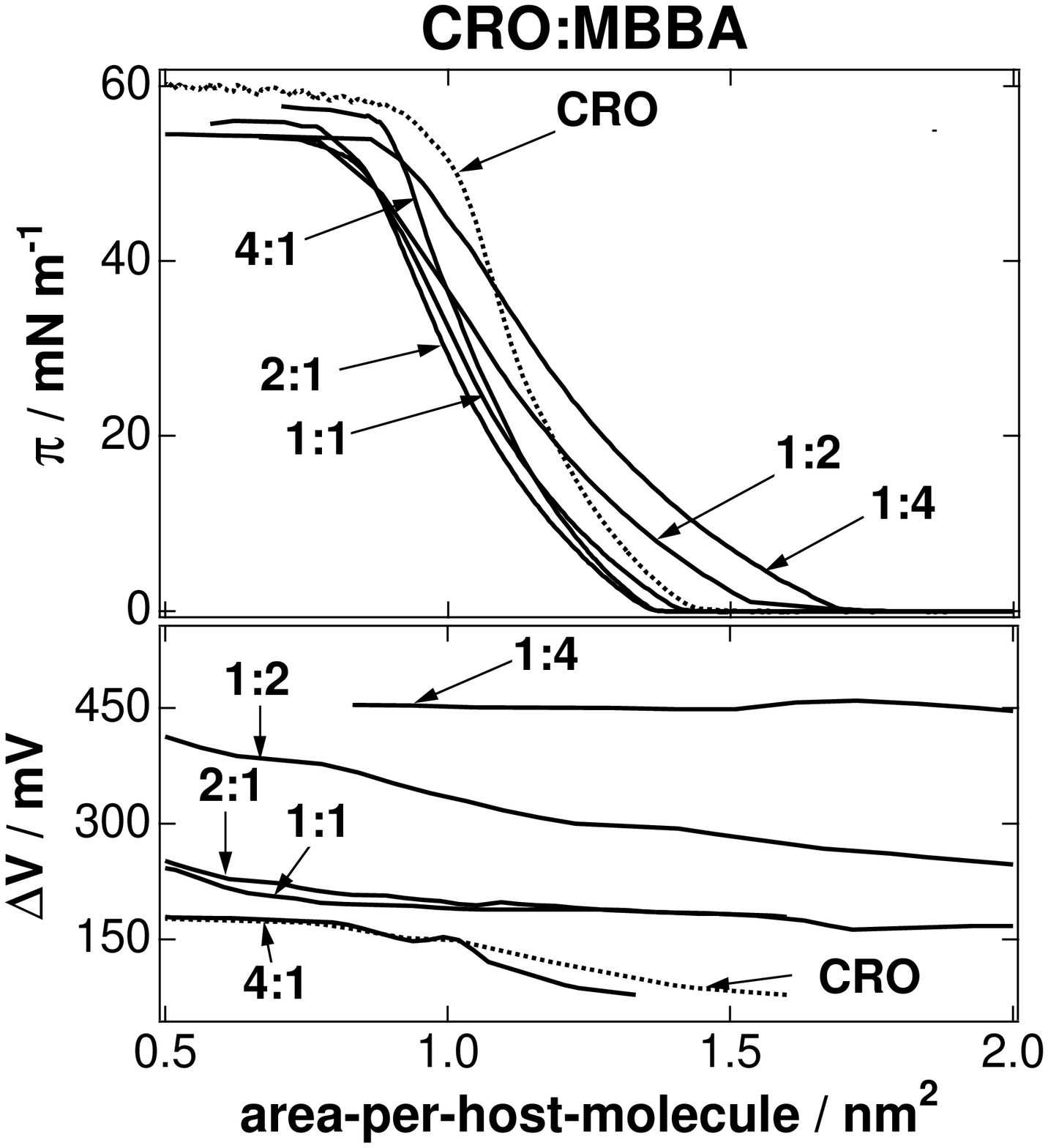, width=0.495\textwidth}
\hfill
\epsfig{file=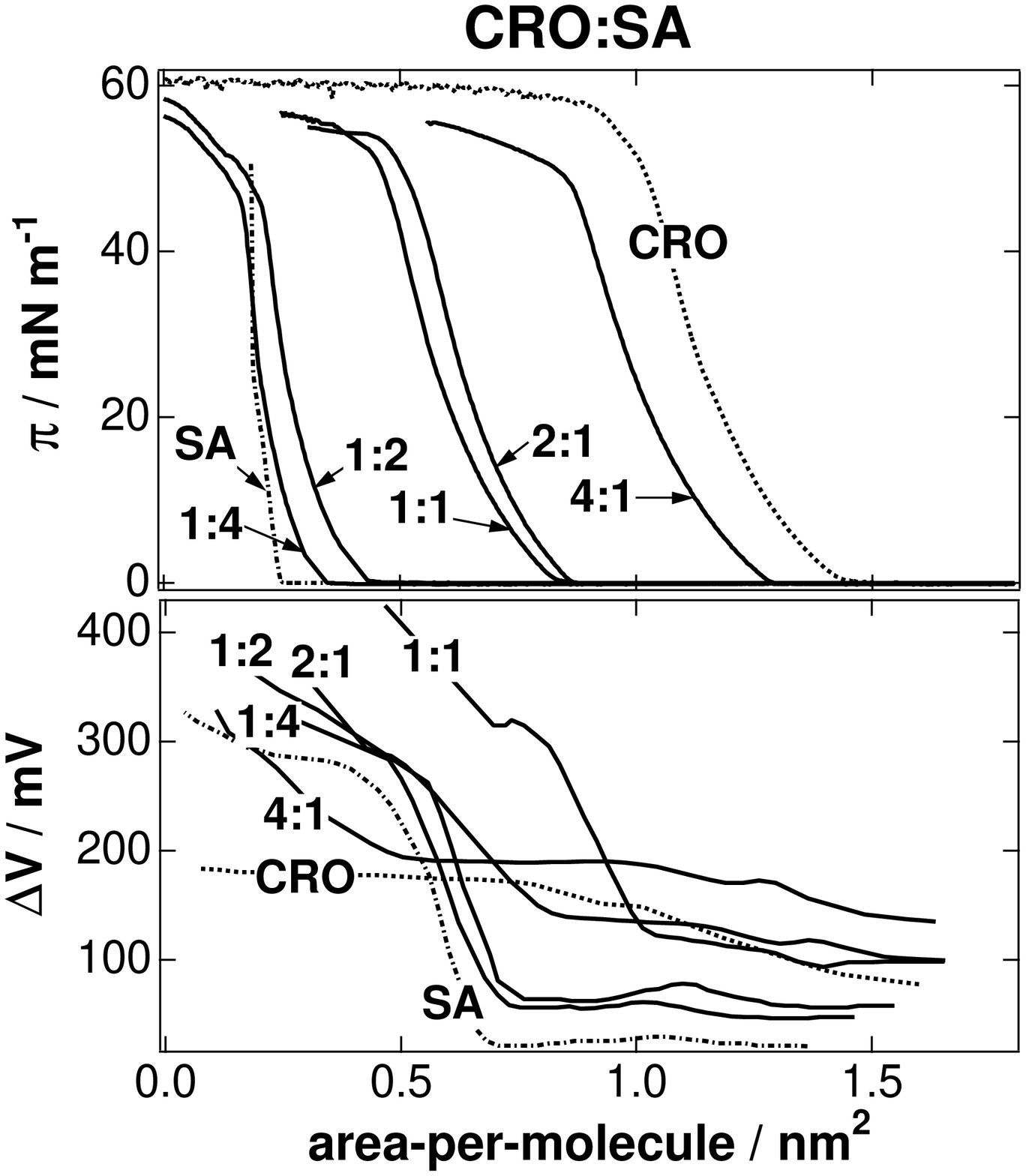, width=0.495\textwidth}
\caption{
\label{CRO-MBBA-SA}
\small{$\pi/A_{\text{h}}$ (cfr. Equation \protect\ref{eq-Ah})
and $\Delta V/ A_{\text{h}}$ isotherms of CRO-MBBA mixed monolayers 
and $\pi/A$ and $\Delta V/ A$ isotherms of CRO-SA mixed monolayers.
The proportions of CRO to the guest compounds are indicated. 
}}
\end{center}
\end{figure}

\begin{figure}[t]
\begin{center}
\epsfig{file=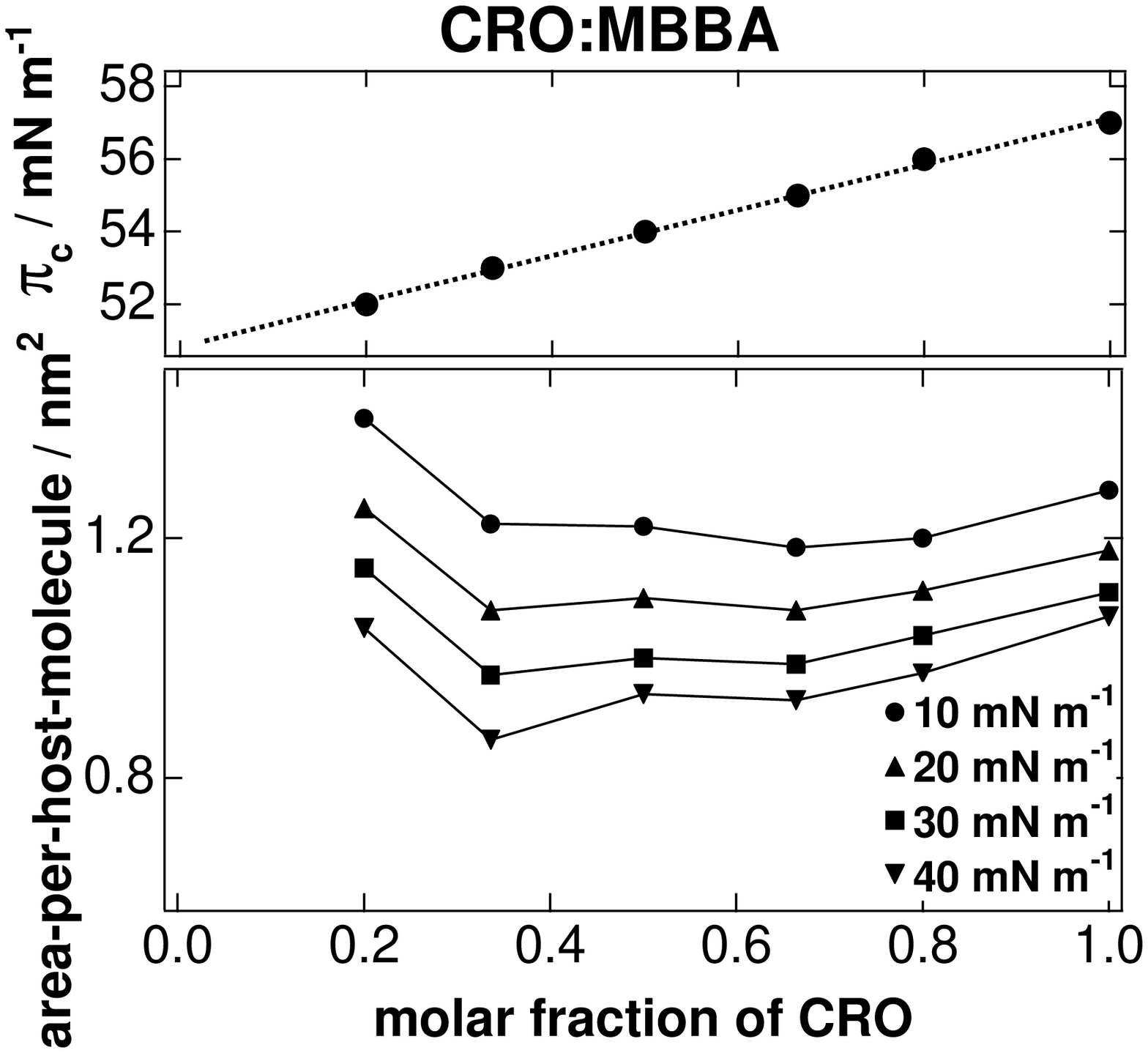, width=0.48\textwidth}
\hfill
\epsfig{file=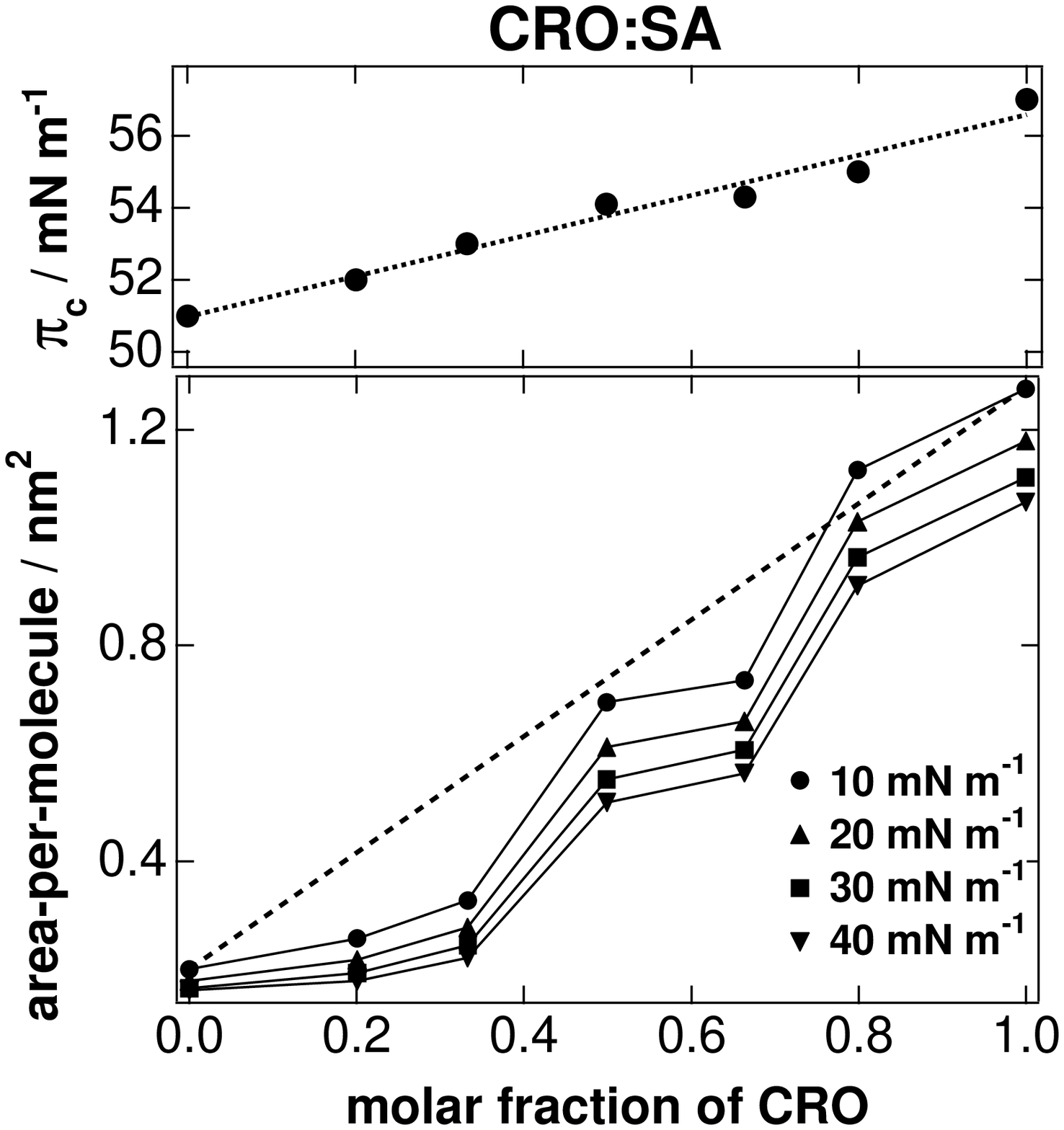, width=0.48\textwidth}
\caption{
\label{area-coll}
\small{Extrapolated $A_{\text{h}}$ (CRO-MBBA) and $A$ (CRO-SA) at 
different surface pressures, and collapse pressures, as a function of
the molar ratio of CRO for the different mixed monolayers.}
}
\end{center}
\end{figure}

We prepared mixed solutions of CRO-MBBA and CRO-SA with different molar 
ratios and studied the behavior of the mixed monolayers at the a/w 
interface.
Since MBBA does not form monolayers\cite{FazKomLagMobXX}, in the study of 
the CRO-MBBA isotherms it is more useful to consider the 
\textit{area-per-host-molecule}, $A_{\text{h}}$, defined as:
\begin{equation}
A_{\text{h}} = \frac{\text{area of the trough}}
{\text{number of host molecules}} = A \frac{\text{total number of molecules}}
{\text{number of host molecules}}.
\label{eq-Ah}
\end{equation}
Instead, for the mixtures CRO-SA we will consider the 
area-per-molecule because both compounds are amphiphilic.
In Figure \ref{CRO-MBBA-SA} the $\pi/A$ and $\Delta V/ A$ isotherms 
of CRO-MBBA and CRO-SA are shown respectively.
In Figure \ref{area-coll} the extrapolated $A_{\text{h}}$ (CRO-MBBA) 
and $A$ (CRO-SA) at different surface pressures, as well as the collapse 
pressures, of the mixed monolayers are shown as a function of the 
molar ratio of CRO in the mixtures.

Let us examine the CRO-SA mixed monolayers first, which are both 
amphiphilic monolayer-forming compounds.
The properties of a monolayer in which the two components are 
immiscible will reflect those of the two separate single component 
films\cite{Gaines}: the area occupied by the mixed film, $A_{\text{mix}}$,
will be the sum of the areas of the two separate films,  
$A_{\text{CRO}}$ and $A_{\text{SA}}$, weighted with the molar 
fraction of the two components, $N_{\text{CRO}}$ and $N_{\text{SA}}$:
\begin{equation}
A_{\text{mix}} = N_{\text{CRO}} A_{\text{CRO}} + N_{\text{SA}}
A_{\text{SA}}.
\label{no-int}
\end{equation}
Thus, in non interacting mixed monolayers $A_{\text{mix}}$ should follow 
a linear relation with $N_{\text{CRO}}$ or $N_{\text{SA}}$.
Any deviation from this linear behavior (also called \textit{line of 
no interaction}) is evidence of interactions in the mixed monolayer, 
that could be attractive or repulsive, and make the mixed films more 
or less stable than the single compounds ones.

The analogs of miscibility or immiscibility of mixed monolayers in 
which one of the two components is not amphiphilic, are 
\textit{cooperative} or \textit{additive incorporation}.
In the first case, the guest molecules interact with the host ones
and there are deviations from the linear relation of Equation 
\ref{no-int}.
Since the guest compound does not form monolayers, Equation 
\ref{no-int} takes in this case the following form:
\begin{equation}
A_{\text{mix}} = N_{\text{h}} A_{\text{h}},
\end{equation}
where $N_{\text{h}}$ is the molar fraction of the host compound and 
$A_{\text{h}}$ is the area-per-host-molecule defined in Equation 
\ref{eq-Ah}.

On increasing SA surface concentration CRO-SA isotherms (Figure 
\ref{CRO-MBBA-SA}) shift towards smaller area-per-molecule and their shape,
as well as the behavior of the surface-potential, resemble more and more
that of stearic acid.
From Figure \ref{area-coll} is evident that 
mixed monolayers CRO-SA with a low molar fraction of SA are immiscible 
as $A_{\text{mix}}$ lies on the line of no interaction or above it 
(repulsive interactions).
Increasing the molar ratio of SA the interactions become attractive 
and the monolayers become more compressed and stable.

Instead, CRO-MBBA $\pi/A_{\text{h}}$ isotherms are not very much 
influenced by the addition of MBBA (Figure \ref{CRO-MBBA-SA}).
Only for high concentrations of MBBA (1:2 and 1:4) the films become 
more compressed.
On the contrary, the surface-potential $\Delta V$ is considerably 
influenced and increases as the molar fraction of MBBA increases,
due to MBBA's CH$_{3}$ groups that contribute positively to the 
surface potential ($\mu_{\text{CH}_{3}}$ = 0.351\,D).
From Figure \ref{area-coll} is evident that MBBA is additively 
incorporated into CRO's monolayers: $A_{\text{h}}$ shows almost no 
change on increasing the molar fraction of the liquid crystal:
MBBA molecules penetrate into the CRO's cavity with their OCH$_{3}$ 
groups oriented towards the polar headgroup (COOH) of the host 
molecules\cite{FazKomLagMobXX}.
Only for very high molar fractions of MBBA (1:4) the films seem to be 
unstable as $A_{\text{h}}$ increases: MBBA molecules in excess are 
squeezed out the monolayer and the isotherm is destabilized. 

\subsection{LB monolayers: ellipsometric and spectroscopic 
characterization}

LB films were prepared depositing 3 or 5 monolayers onto 
chro- mium-sputtered plates for ellipsometric measurements and 
1 single monolayer onto quartz plates for spectroscopic measurements.
The deposition surface-pressure was 25\,mN\,m$^{-1}$.
For the first monolayers on both substrates we always obtained a 
transfer ratio, TR, larger than unity: TR $\approx$ 1.2, which seems 
to be typical of calixarenes (see for instance \cite{MoeDutAro94}).
The other 2 or 4 monolayers were transferred with a TR very close to 
unity.

The results of the ellipsometric measurements are listed in  
in Table \ref{depos}.
\begin{table}
\begin{center}
\caption{
\label{depos}
\small{Monolayer thickness, $t$, as estimated from ellipsometric 
measurements.}
} 
\begin{tabular}{|c|c|}
\hline
\textbf{CRO} & $t$ [nm] \\
\hline
 & 1.3 \\
\hline
\end{tabular}
\begin{tabular}{|c|c|}
\hline
\textbf{CRO-MBBA} &  $t$ [nm]  \\
\hline
4:1 & 1.5 \\
\hline
2:1 & 1.7 \\
\hline
1:1 & 1.8 \\
\hline
1:2 & 2.0 \\
\hline
1:4 & 2.0 \\
\hline
\end{tabular}
\begin{tabular}{|c|c|}
\hline
\textbf{CRO-SA} &  $t$ [nm]  \\
\hline
4:1 & 1.0 \\
\hline
2:1 & 0.8 \\
\hline
1:1 & 1.0 \\
\hline
1:2 & 0.7 \\
\hline
1:4 & 0.7 \\
\hline
\end{tabular}
\end{center}
\end{table}
In the case of pure CRO, from a comparison between the measured monolayer 
thickness and the length of the hydrocarbon chains (1.65\,nm) it was 
possible to estimate that the chains were tilted of an angle 
$\theta = 51 (\pm 5$) degrees, which is in agreement with the previous 
tilt angle estimation by surface-potential measurements.
In the case of CRO-MBBA LB films, the monolayer thickness are larger 
than that of single CRO monolayers and increase with increasing  
MBBA concentration, which indicates that MBBA is present in the LB 
films and that deposition does not cause molecular rearrangement or 
collapse in the monolayers.
On the contrary, the small thickness measured in the case of CRO-SA LB 
films, confirm the monolayer instability already observed at the a/w 
interface.

\begin{figure}[h!]
\begin{center}
\epsfig{file=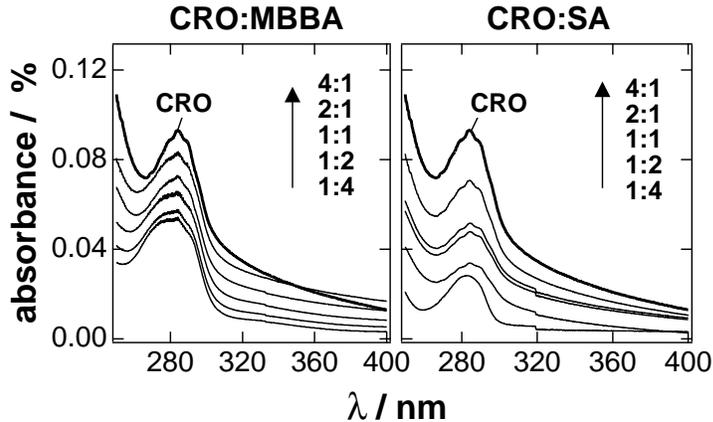, width=0.75\textwidth}
\caption{
\label{spectra}
\small{UV-spectra of the mixed monolayers deposited onto quartz plates.}
}
\end{center}
\end{figure}

\begin{figure}[t]	
\begin{minipage}{\textwidth}
\parbox[b]{0.55\textwidth}{
\epsfig{file=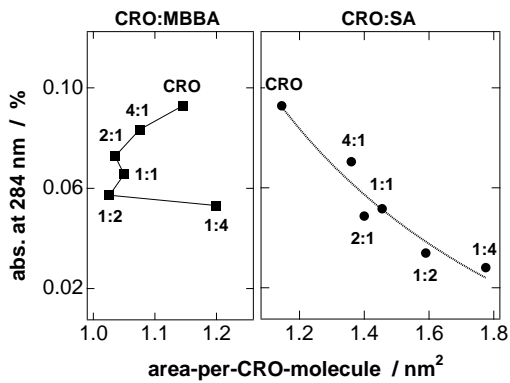, width=0.53\textwidth}
} \hfill
\parbox[b]{0.44\textwidth}{
\caption{
\label{uv-calc}
\small{Intensity of the UV spectra at 284\,nm as a function of the 
area per molecule.
The markers are the experimental values.
The dotted line in the CRO:SA graph is a fit of the experimental data 
to a function invertionally proportional to the area-per-CRO-molecule.
}
}
}
\end{minipage}
\end{figure}

The results of the spectroscopic measurements are shown in Figure 
\ref{spectra}.
Both CRO and MBBA absorb at 284\,nm.
The height of the peaks in the CRO-MBBA spectra is not only due to a 
change in surface density of the CRO molecules.
If so, the intensity of the peak at 284\,nm should be proportional to the 
surface density of CRO molecules, i.e. invertionally proportional to 
area-per-CRO-molecule.
Figure \ref{uv-calc} shows the absorption at 284\,nm as a function of 
the area-per-CRO-molecule in the various CRO-MBBA and CRO-SA mixed 
LB films.
The figure reveals that MBBA is present in the deposited monolayers
and contributes to the UV-spectrum.
On the other hands, since SA does not contribute to the spectrum at those 
wavelengths, we expected the intensity of the peak to depend only on 
the surface density of CRO molecules, i.e. to be invertionally 
proportional to the area occupied by the CRO molecules.

\section{Conclusions}
\noindent
A comparative study of mixed CRO-MBBA and CRO-SA films has been 
carried out in order to achieve a better understanding of the  
interactions in guest-host monolayers.

In CRO-SA mixed layers interactions occur that cause the films not to 
be always stable. 
Repulsive interactions occur for low molar fractions of SA that
destabilize the monolayers at the a/w interface and cause collapse 
and/or molecular rearrangement during transfer of the films onto solid 
substrates.
Attractive interactions occur for high molar fractions of SA and the 
monolayers at the a/w interface are stable.
However, the transfer seems to cause destabilization of the films.

On the other hand, mixed CRO-MBBA films are stable and easily 
transferable as long as the molar fraction of MBBA is not too large 
(in which case the MBBA molecules in excess are squeezed out of the 
monolayer during compression).
MBBA results to be additively incorporated into CRO monolayers.

The mechanism of additive incorporation of MBBA in CRO films resembles 
that of alignment of LCs on surfactants.
As MBBA penetrates into the cavity of CRO apparently without any specific 
interaction, so LC molecules penetrate into the chain region of the 
surfactant and align the LC bulk: the alignment can be very stable 
even without strong interactions between the LC and the aligning layer.

\section{Acknowledgments}
F. Nannelli is grateful to the C.M. Lerici Foundation (Institute of 
Italian Culture, Stockholm, Sweden) and V. S. U. Fazio is grateful 
to the European TMR Programme (contract number ERBFMBICT9830 23) and to
the W. \& M. Lundgrens Foundation (G\"oteborg, Sweden) for financial 
support.
L. Komitov acknowledges the financial support of the Japanese Society for 
Promotion of Sciences.

\bibliography{journal2,calix}
\end{document}